\documentclass[
    ,final            % use final for the camera ready runs
  ]
  {aipproc}

\layoutstyle{8x11single}

\begin{document}

\title{Nuclear and gravitational energies in stars}

\classification{97.10.-q, 97.60.-s, 95.30.Sf, 21.10.-k }
\keywords      {Stars, nuclear reaction, gravity}

\author{Georges Meynet}{
  address={Astronomical Observatory of Geneva University }
}

\author{Thierry Courvoisier}{
  address={ISDC, Astronomical Observatory of Geneva University}
}

\author{Sylvia Ekstr\" om}{
  address={Astronomical Observatory of Geneva University }
}

\begin{abstract}
The force that governs the evolution of stars is gravity. Indeed this force drives star formation, imposes thermal and density gradients into
stars at hydrostatic equilibrium and finally plays the key role in the last phases of their evolution. 
Nuclear power in stars governs their lifetimes and of course the stellar nucleosynthesis. 
The nuclear reactions are at the heart of the changes of composition of the baryonic matter
in the Universe. This change of composition, in its turn, has profound consequences on the evolution of stars and galaxies. 

The energy extracted from the gravitational, respectively nuclear reservoirs during the lifetimes of stars of different masses are estimated. 
It is shown that low and intermediate mass stars (M $<$ 8 M$_\odot$) extract roughly 90 times more energy from their
nuclear reservoir than from their gravitational one, while massive stars (M $>$ 8 M$_\odot$), which explode in a supernova explosion, extract more than 5 times
more energy from the gravitational reservoir than from the nuclear one.
We conclude by discussing a few important nuclear reactions and their link to topical astrophysical questions.
\end{abstract}

\maketitle

%%%%%%%%%%%%%%%%%%%%%%%%%%%%%%%%%%%%%%%%%%%%
%% MAINMATTER
%%%%%%%%%%%%%%%%%%%%%%%%%%%%%%%%%%%%%%%%%%%%

\section{The leading role of gravity}

In this paragraph we mention a few textbook results which are at the heart of the understanding of the evolution of stars. More details can be
obtained for instance in the following books: \citet{Kipp}, \citet{Maeder 2009}.
Stars continually lose energy from their surface. This loss at the surface is a consequence
of the equilibrium of two forces in the interior of stars. On one side, there is the gravity which holds together the gas. Alone this force
would make matter fall to the center in a free fall timescale proportional to $1/\sqrt{G \overline{\rho}}$, where $G$ is the gravitational constant and $\overline{\rho}$, the mean density of the object. On the other side, there is the force resulting from pressure gradients which tends to inflate the star. This last force is linked to the random movement of the gas particles and to the pressure due to photons (radiative pressure)\footnote{For a star like the sun, the radiation pressure is negligible and the whole weight is supported by the gas pressure. In contrast for more and more massive stars, radiation pressure becomes more and more important. Typically, in a 60 M$_\odot$ star, during the Main-Sequence phase, the radiation pressure contributes to about 30\%
of the total pressure, while in the 1 M$_\odot$, it contributes only to 5 ten thousandths to the total pressure in most of the interior.}. It implies that the material must be hot and therefore must emit radiation. Thus the luminosity of a star is a necessary consequence of its hydrostatic equilibrium.

An expression for the luminosity can be obtained without
reference to any specific source of energy. Using the hydrostatic equilibrium equation, the radiative transfer equation and an equation of state (see e.g.
\citet{Maeder 2009}), it can be shown that the luminosity is roughly proportional to $\mu^4 M^3 /\kappa$, where $\mu$ is the mean molecular weight
({\it i.e.} the mass per free particle in units of an atomic mass unit) and $\kappa$ the opacity, these two quantities being representative for the whole
material making the star. Thus  the luminosity of a star is determined from global properties of the star and not  by the specific mechanisms
which produce the energy. These mechanisms adapt themselves in order to compensate for the losses of energy at the surface of the star.
Note that beside the luminosity, many other properties of stars, such as the central pressure, temperature or density
can be derived from the hydrostatic equilibrium and an equation of state.
Very interestingly also a natural scale for the mass of stars can be obtained, expressed only in terms of fundamental constants of physics (\citet{Chandraskhar 1984}).

Where do the nuclear reactions get involved in this picture?  They are involved when one tries to estimate the lifetimes of stars. In absence of any nuclear reactions, stars would evolve in much shorter timescales. This timescale would be the Kelvin-Helmholtz timescale which can be simply expressed
as half the ratio of the gravitational energy to the luminosity of the star, {\it i.e} $G M^2/(2 R L)$, where $M$, $R$ and $L$ are respectively the mass, radius and luminosity of the star. 
This is of course a rough estimate. But typically for a star like the sun
it amounts to a value of about 30 My, while for a 20 M$_\odot$, it would be of the order of 100 000 y. Actually, this timescale corresponds to the duration
of the slow contraction from the Hayashi line to the ZAMS, {\it i.e.} of a phase during which gravity is quasi the only energy source in the star (quasi because some energy may be produced by nuclear reactions involving very light and fragile chemical species as deuterium for instance).

Let us explain in more details the meaning of the Kelvin-Helmholtz (KH) timescale, starting from an equilibrium situation between gravity and the pressure gradient. Without any heating mechanism, the thermal gradient
inside a star would on the long term disappear and the temperature will decrease, like when one switches off the source of heat below a boiling water pot. This however cannot happen in stars because of the gravity. Indeed, when the pressure gradient decreases, gravity is no longer perfectly counterbalanced and
thus the gas contracts. The contraction is slow because pressure gradients are present and thus this contraction does not occur in a
free fall timescale as it would if gravity were acting alone. While contracting, the gas warms up, building up a new pressure gradient, able, at least temporarily, to again counterbalance gravity. The energy which serves to heat the material comes from gravity. Actually, the Virial theorem (see e.g. \citet{Maeder 2009}) 
tells us that only about half of the gravitational energy released by the contraction is used to
warm up the material (and thus used to reach a new hydrostatic equilibrium), the other half is radiated away. 
This explains the factor 1/2 in the expression of the Kelvin-Helmholtz
timescale. The quantity $GM^2/2R$  can be interpreted as the thermal energy which has been accumulated inside the star since the beginning of its contraction, a similar energy having been radiated away by the star. 
%Note that the initial radius was so great with respect to
%the present radius of the star  that its value is of no importance and can be considered as infinite without any  impact on the KH timescale.

Interestingly, this KH timescale was one of the first quantitative and physically motivated estimates of the age of the Sun (see a modern account of this story in \citet{Stinner}). This estimate received a strong support by the fact that
the age of the earth deduced from its past thermal history gave an age compatible with the KH lifetime of the Sun! 
However geologists at that time (end of the nineteenth century), or the tenants
of the Darwin theory for the origin of the species required for valleys to be formed, or for species to evolve, much longer timescales.
So there was a paradox! Various processes, observed on our planet, indicated ages for the earth orders of magnitudes greater than the age of the Sun. This was particularly embarrassing since erosion, as well as life need the solar energy to occur!

Gravitational energy reservoir is large enough for explaining the luminosity of stars, but it gives too short timescales.
So there comes the need for another source of energy than gravity, and this is where nuclear reactions come into play (see Chapter 3 in \citet{Longair} for a very nice historical account of this
discovery). A very rough expression for the Main-Sequence lifetime of a star is given by the expression
$X q M 0.007 c^2/L$
where, $X$ is the mass fraction of hydrogen (typically 0.7), $q$ the fraction of the total mass of the star where core H-burning occurs (about 10\% for a 1 M$_\odot$ star), $c$, the velocity of light. This expression is based on the fact that
every time a unit mass of hydrogen is transformed into helium, 7 thousandths of that mass is transformed into energy. For the Sun, this expression gives a lifetime of the order of 10 Gy thus
much longer than the KH timescale.

Nuclear energy allows a pressure gradient to be maintained without the need for the star to contract too much. 
Still some contraction  of the central regions occurs because the fuel diminishes.
When a fuel has been completely consumed in the core,
the core has to tap into the gravitational reservoir and contracts until physical conditions for
new nuclear reactions to occur are realized or until the star arrives at the end point of its evolution (see more below).
During the core H and He-nuclear burning phases, because contraction is slowed down,
the central temperature
varies less than during the phase when no nuclear burning occurs in the core. In the advanced phases of massive stars, neutrino emission may remove quantities of energies equal to those
produced by the nuclear reactions, in that case gravity becomes the main source for compensating the energy radiated away at the surface.
\begin{figure*}
  \includegraphics[height=.33\textheight]{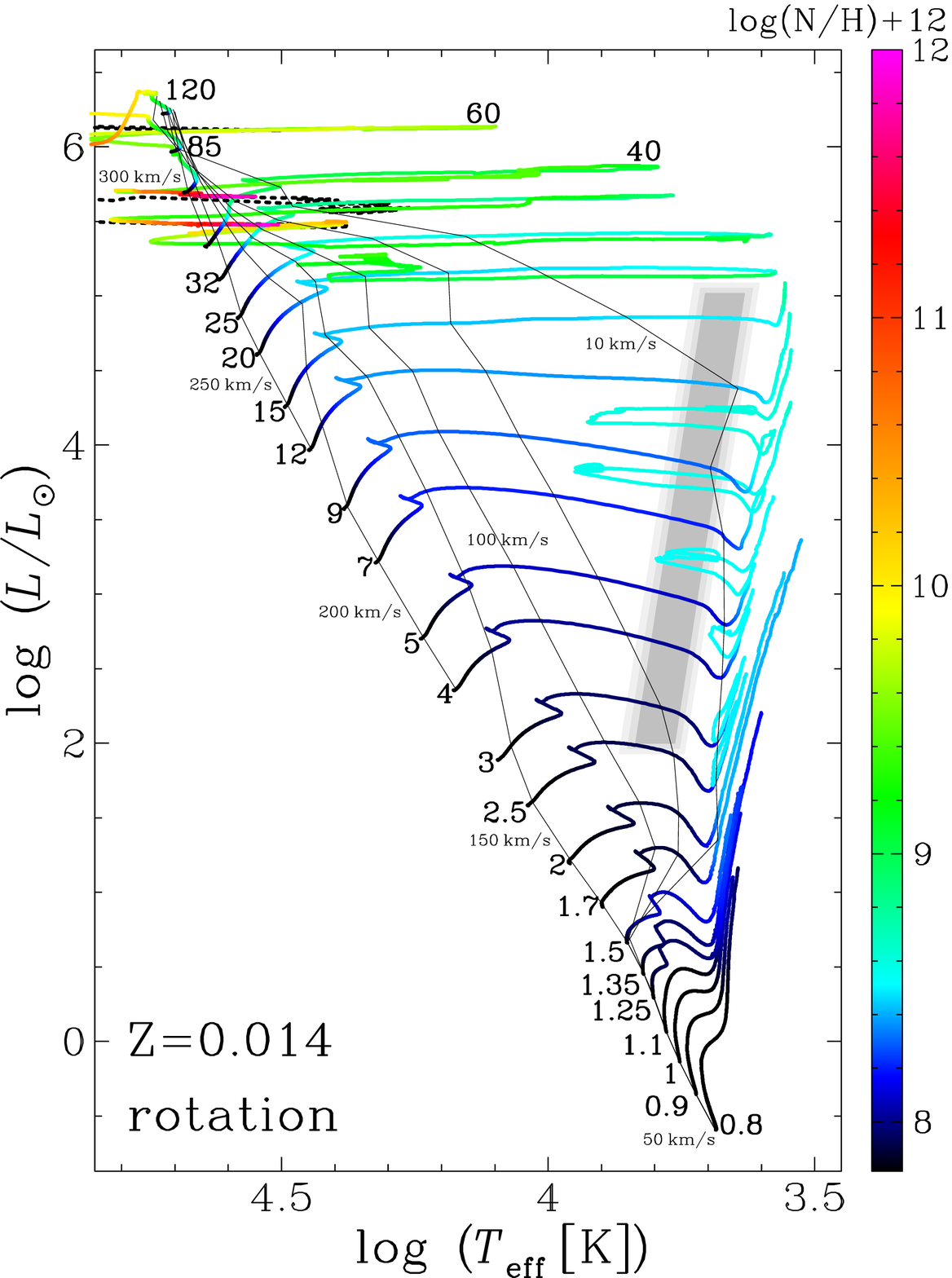}  \includegraphics[height=.33\textheight]{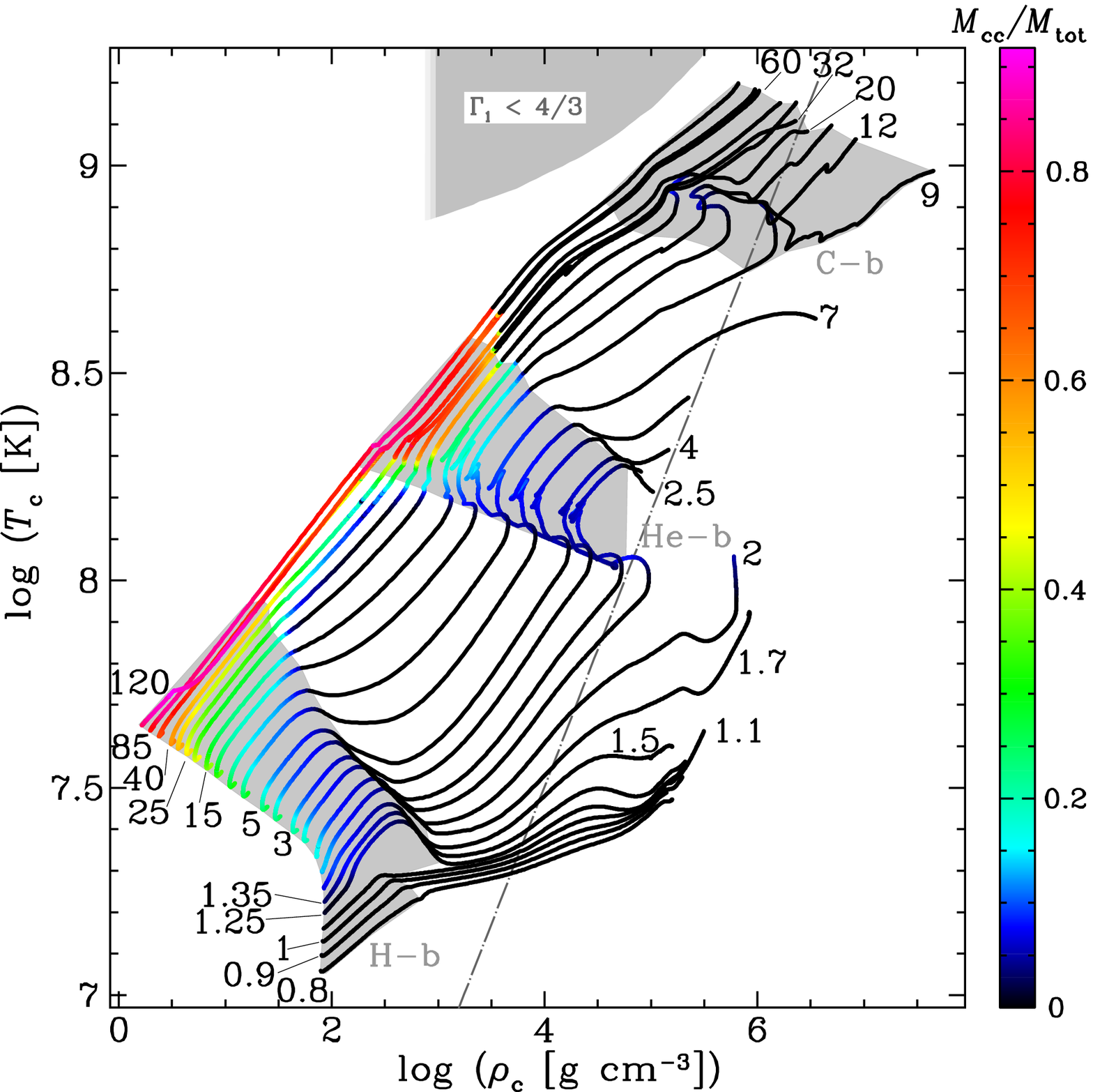}
  \caption{{\it Left panel:} Evolutionary tracks in the theoretical Hertzsprung-Russel diagram for stars with masses between 0.8 and 120 M$_\odot$. The metallicity is solar ($Z$=0.014) and the stars start their evolution on the ZAMS with a surface equatorial velocity equal to 40\% the critical velocity. The critical velocity is the velocity that the star should have for the centrifugal force at the equator to be equal to the gravity. The colors indicate the level of surface nitrogen enrichment at the surface in terms of the nitrogen to hydrogen ratio (in number). The grey shaded area 
  shows the Cepheid instability strip. This figure is taken from \citet{Ekst}). {\it Right panel:} Evolution of the same models as in the left panel in the plane central temperature versus central density. The colors indicate the size of the convective core in fraction of the total mass of the star. The grey shaded areas shows the part of the tracks where nuclear burning occur during the H-, He- and C-burning phases.}
\end{figure*}

The evolution of a star is a continuous contraction at least of the central parts and very high densities are reached when the stars arrives at the end of its lifetime.
Most stars, except may be the most massive ones (see below), or those forming a black hole, will lock part of their core in an object sustained by
a degenerate gas of electrons (white dwarfs for initial masses below about 8 M$_\odot$) or of neutrons (for initial masses above 8 M$_\odot$, although not all stars will end as neutron stars in this mass range, some will give birth to black hole or are completely disrupted). When the gas is completely degenerate, the pressure depends only on the density and
thus the pressure gradient, needed to sustain the mass against gravity, is obtained through a density gradient. Now, such a density gradient has no
possibility to evolve being locked by gravity. This is in contrast with a temperature gradient which always triggers an energy flux and thus implies an evolution. Thus a star sustained by a degenerate gas no longer evolves. 

Actually this is not exactly true because in a white dwarf, the ions are not degenerate and thus can cool. This makes the cooling sequence of white dwarfs
(see e.g. \cite{Isern}).
Moreover, in case a companion is sufficiently near the WD, some mass transfer may occur. This induces high temperatures in the accreted material which  produces a nova, or in some circumstances causes the collapse of the WD. This collapse induces nuclear reactions in highly degenerate regimes which completely blow up the star
(type Ia supernovae). In a similar way, neutron stars and black hole do no longer evolve in the sense that their state does not imply any radiation (at least  in usual terms).

\section{nuclear versus gravitational energy in stars}

Just above, we saw the respective roles of the gravitational and nuclear power in stars. 
Let us now estimate the amount of energy released by gravitational contraction and by nuclear reaction during the whole lifetimes of stars of various
initial masses. 
As we shall see this leads to a marked difference between the low and intermediate mass stars on one side  and the massive stars on the other.
A first estimate of these energies has been presented in \cite{Courvoisier2013}, with interesting considerations about the source of energy to accelerate the cosmic rays.
 
\subsection{The case of low and intermediate mass stars (M $<$ 8 M$_\odot$)}

Let us begin by estimating the energy extracted from gravity by a 1 M$_\odot$ star.
This energy can be obtained simply by computing the binding energy of the remnant, which is a white dwarf
of about 0.5 M$_\odot$ (M$_\odot \approx 2 \times 10^{33}$ g) whose radius is about 0.001 R$_\odot$ (R$_\odot \approx 7 \times 10^{10}$ cm). One obtains 
$E_{\rm grav}=G \times (0.5 \times M_\odot)^2/(0.001 \times R_\odot)=10^{50}$ ergs or
6 $\times$ 10$^{61}$ eV.

Let us now compute the energy extracted from the nuclear reservoir. The white dwarf is made mostly of carbon and oxygen. The binding energy per nucleon in oxygen  is about 8 MeV per nucleon (the binding energy of carbon nuclei is not very different). Thus the total energy
coming from the nuclear reservoir is, $M_p$ being the mass of a nucleon,
$$E_{\rm nucl}={0.5 \times M_\odot \over M_p} 8= 500 \times 10^{61} {\rm eV}=5 \times 10^{57} {\rm MeV}.$$
One sees that for a 1 M$_\odot$ star, the total energy from
nuclear reactions is about 500/6=83 times larger than the energy extracted
from gravity. If we divide the energy extracted from
nuclear reactions by the solar luminosity one obtains 66 Gy. This is greater than
the solar lifetime because the luminosity is not constant during the whole
lifetime. We also overestimate a little here the nuclear energy released
because not all the mass of the white dwarf will be composed of carbon and oxygen.

%\subsection{Variation as a function of the initial mass for $E_{\rm grav}$ and $E_{\rm nucl}$ for low and intermediate mass stars}

Let us now consider how these energies vary as a function of the initial mass.
For stars evolving into WD, the WD mass can vary between about 0.5 and 1.4
M$_\odot$, the radius of the white dwarfs varies as $M^{-1/3}$. The
binding energy of a WD  with mass equal to $M_{\rm WD}$ is,
using the result above for a 0.5 M$_\odot$ WD:
$$E_{\rm grav}={(M_{\rm WD}/0.5)^2\over(M_{\rm WD}/0.5)^{-1/3}}6 \times 10^{61}{\rm eV}=\left({M_{\rm WD}\over 0.5}\right)^{7/3}6 \times 10^{61}{\rm eV}.$$

For $E_{\rm nucl}$, we have

$$E_{\rm nucl}={M_{\rm WD} \over 0.5} 500 \times 10^{61} {\rm eV}.$$

The ratio between the gravitational and nuclear energy becomes
$${E_{\rm grav} \over E_{\rm nucl}}  =\left({M_{\rm WD} \over 0.5}\right)^{4/3} 0.012.$$

%\subsection{Nuclear versus gravitational energy for a population of low and intermediate stars}

The mass of the white dwarf can be related to the initial mass, M (in M$_\odot$), through relations
obtained from ``observations" (between brackets because this kind of
observation involves a lot of theory, see \citet{Catalan et al. 2009}). For initial masses below 2.7 M$_\odot$, one has
$$M_{\rm WD}=0.096 \times M+0.429.$$
For initial masses above 2.7 M$_\odot$ and below 8 M$_\odot$, one has
$$M_{\rm WD}=0.137 \times M+0.318.$$
Using the Salpeter Initial Mass Function, the number of stars with masses between $M$ and $M$+d$M$ is given by d$N$/d$M$=C $M^{-2.35}$, where C is a constant.
The energy released from gravity by stars with initial masses
between 0.9 and 2.7 M$_\odot$ per star in a stellar generation is therefore
$$<E_{\rm grav} (0.9-2.7)>=6 \times 10^{61} {\int_{0.9}^{2.7} M^{-2.35} \left({0.096 \times M+0.429\over 0.5} \right)^{7/3} {\rm d}M \over \int_{0.01}^{120} M^{-2.35}  {\rm d}M}=6 \times 10^{61}{0.4997 \over 371}\sim0.01 \times 10^{61} {\rm eV}.$$
It is worthwhile to make a few remarks here
\begin{enumerate}
\item
  This is the energy released when all stars have terminated their evolution. We chose the value of 0.9 M$_\odot$
  for the lower bound of the masses that can contribute to the release of gravitational energy since the lifetime of these stars is just of the same order as the age of the Universe.
  This estimate does not include the energy released by type Ia SNe. It should be counted in $E_{\rm nucl}$, since this type of SNe are triggered by the nuclear energy.
\item
For obtaining the energy released from gravity by stars with
masses between 0.9 and 2.7 M$_\odot$, when N stars are born with
masses between 0.01 and 120 M$_\odot$, it suffices to multiply the
number above by N.
\item 
If one would like to have the energy released extracted from gravity by stars with
masses between 0.9 and 2.7 M$_\odot$ per solar mass locked into star in one generation,
the denominator should be the integral of $M^{-1.35}$ over 0.01 and 120 M$_\odot$.
%\item Note that here we used a Salpeter IMF.
\end{enumerate}
In a similar way, the energy released from gravity by stars with initial masses between 2.7 and
8 M$_\odot$ per star in a given stellar generation is
$$<E_{\rm grav} (2.7-8)>=6 \times 10^{61} {\int_{2.7}^{8} M^{-2.35} \left({0.137 \times M+0.318\over 0.5} \right)^{7/3} {\rm d}M \over \int_{0.01}^{120} M^{-2.35}  {\rm d}M}=6 \times 10^{61}{0.3034 \over 371}\sim0.005 \times 10^{61} {\rm eV}.$$
Thus the order of magnitude is the same as for the mass range between 0.9 and 2.7 M$_\odot$.

We compute now the energy extracted from the nuclear reservoir, again distinguishing
the mass range between 0.9 and 2.7 M$_\odot$ and the one between 2.7 and 8 M$_\odot$. One obtains
respectively the two expressions below:
$$<E_{\rm nucl} (0.9-2.7)>=500 \times 10^{61} {\int_{0.9}^{2.7} M^{-2.35} \left({0.096 \times M+0.429\over 0.5} \right) {\rm d}M \over \int_{0.01}^{120} M^{-2.35}  {\rm d}M}=500 \times 10^{61}{0.7482 \over 371}\sim 1.01 \times 10^{61} {\rm eV}.$$
$$<E_{\rm nucl} (2.7-8)>=500 \times 10^{61} {\int_{2.7}^{8} M^{-2.35} \left({0.137 \times M+0.318\over 0.5} \right) {\rm d}M \over \int_{0.01}^{120} M^{-2.35}  {\rm d}M}=500 \times 10^{61}{0.2697 \over 371}\sim0.36 \times 10^{61} {\rm eV}.$$

\subsection{The case of massive stars (M $>$ 8 M$_\odot$)}

For a 20 M$_\odot$ star, the remnant will be a neutron star of
about 1.5 M$_\odot$, of radius about 10 km. Thus one has
$$
E_{\rm grav}={G \times (1.5 \times M_\odot)^2 \over 1.4 \times 10^{-5}R_\odot}\approx 38200\times 10^{61} {\rm eV}\approx 6 \times 10^{53} {\rm ergs}.
$$
The energy extracted from gravity by a 20 M$_\odot$ is more than 6000 times larger than for a 1 M$_\odot$ star.

The mass of neutron stars may vary between about 1.5 and 2.5 M$_\odot$. The radius varies as $M^{-1/3}$. For a neutron star of mass $M_{\rm NS}$, one has
$$
E_{\rm grav}=\left({M_{\rm NS} \over 1.5} \right)^{7/3} 38200\times 10^{61}{\rm eV}.
$$
Typically, for $M_{\rm NS}$=2.5 M$_\odot$, the energy extracted from gravity will be 
$(2.5/1.5)^{7/3}$ larger than for
the 1.5 M$_\odot$, that means a factor 3.3 larger. 

The binding energy per nucleon of iron is 8.8 MeV. The size of the iron core in
a 20 solar mass star is of the order of 1 M$_\odot$, the CO layers, $M_{\rm CO}$, around the iron core have a mass of about 3 M$_\odot$ and the He layers  (6.6 MeV per nucleon), $M_{\rm He}$, around the iron-CO core have a mass around of about 2 M$_\odot$. So the total energy released from nuclear reaction by a 20 M$_\odot$ star is of the order of
$$ 
{M_{\rm Fe}\over m_p} 8.8+{M_{\rm CO}\over m_p} 8.0+{M_{\rm He}\over m_p} 6.6\sim6000\times 10^{61} {\rm eV}=
1\times 10^{53} {\rm ergs}.
$$
For a massive star, the energy extracted from gravity is about 6 times
greater than the energy released by the nuclear reactions. 
In the case of the 20 M$_\odot$, we have that the total mass of the star processed by nuclear burning is M$_{nucl}$=6 M$_\odot$. The averaged binding energy per nucleon in that region is
(8.8+3$\times$8.0+2$\times$6.0)/6= 7.7 MeV. Below, we shall assume that in stars of different initial masses, M$_{nucl}$ is given by the mass interior to the He-rich layers and that
the averaged binding energy per nucleon in that region is equal to 7.7 MeV. 
The mass M$_{nucl}$ scales with the initial
mass as $M_{\rm He} \propto M^{0.92}.$ So the energy extracted from nuclear reactions from a mass M would be
(supposing that 7.7 MeV are extracted per nucleon in M$_{nucl}$):
$$
E_{\rm nucl}=\left({M\over 20}\right)^{0.92} 6000 \times 10^{61} {\rm eV}.
$$
This means that a 9 M$_\odot$ will release an energy about half that of a 20
M$_\odot$, while a 120 M$_\odot$ will release about 5 times the energy of
a 20 M$_\odot$ stars. If we assume that all massive stars evolve into a neutron star of 1.5 M$_\odot$, one has that
$$
{E_{\rm grav} \over E_{\rm nucl}}=\left({20 \over M} \right)^{0.92} \times 6.4.
$$

%\subsection{Variation as a function of the initial mass for $E_{\rm grav}$ and $E_{\rm nucl}$ for massive stars}

The
upper initial mass limit for neutron star formation is not known, probably around 25-30 M$_\odot$ (but likely depends on metallicity, rotation). There is no indication on
how the mass of the neutron star will depend on the initial mass.
%\subsection{Nuclear versus gravitational energy for a population of massive stars}
Let us suppose that all stars between 8 and 30 M$_\odot$ evolve into a 1.5
M$_\odot$ neutron star, while more massive stars evolve into a BH without
any release of energy (likely not a very realistic hypothesis, but the question of what does happen when a black hole is formed is still a matter of discussion. The present hypothesis implies
that the estimate made here is a lower limit to the gravitational energy released by massive stars, but see below).
In that case, the energy released from gravity from
stars with masses between 8 and 120 M$_\odot$ per star in a stellar generation will be
$$<E_{\rm grav} (8-120)>=38200 \times 10^{61} {\int_{8}^{30} M^{-2.35}  {\rm d}M \over \int_{0.01}^{120} M^{-2.35}  {\rm d}M}=38200 \times 10^{61}{0.0372 \over 371}\sim 3.8 \times 10^{61} {\rm eV}.$$
In case one assumes that all stars between 8 and 120 M$_\odot$ evolve
into 1.5 M$_\odot$ neutron stars, the energy released would be 4.5 10$^{61}$ eV\footnote{If we suppose that during a BH formation, 
gravitational energy is released up to the point when the collapsing object reaches a radius equal to the last stable orbit, then about 6\% of the mass of the BH is radiated away.
Assuming that all stars above 30 M$_\odot$ form black holes equal to their initial mass (an extreme hypothesis), then it is easy to estimate the energy released per star in a stellar generation by black hole forming objects, it is equal to 10$^{50}$ ergs per star, or 6.1 10$^{61}$ eV, so less than 2 times the gravitational energy released by neutron star forming objects. This is of course a significative increase but to account for it or not does
not drastically change the main points of this paper. Moreover this is an upper limit since we assume BH with masses equal to the initial mass.}.

The energy released from nuclear reactions from stars with masses between
8 and 120 M$_\odot$  will be
$$<E_{\rm nucl} (8-120)>=6000 \times 10^{61} {\int_{8}^{120} M^{-2.35}  \left({M \over 20}\right)^{0.92} {\rm d}M \over \int_{0.01}^{120} M^{-2.35}  {\rm d}M}=6000 \times 10^{61}{0.042 \over 371}\sim 0.68 \times 10^{61} {\rm eV}.$$

\subsection{Discussion}

\begin{table}
\begin{tabular}{lcccc}
\hline
   \tablehead{1}{l}{b}{Type of Energy}
  & \tablehead{1}{c}{b}{0.9-2.7 M$_\odot$}
  & \tablehead{1}{c}{b}{2.7-8.0 M$_\odot$}
  & \tablehead{1}{c}{b}{8-120 M$_\odot$}   
  & \tablehead{1}{c}{b}{0.9-120 M$_\odot$}\\
\hline
   \tablehead{5}{c}{b}{per star formed between 0.01 and 120 M$_\odot$ in 10$^{61}$ eV} \\
\hline
GRAV  & 0.01 & 0.005 & 3.8    &  3.8\\
NUCL  & 1.01 & 0.36   & 0.7    &  2.1\\
Total   & 1.02 & 0.365 & 4.5    &  5.9\\
\hline
   \tablehead{5}{c}{b}{per solar mass  forming stars with initial masses  between 0.01 and 120 M$_\odot$ in 10$^{61}$ eV} \\
\hline
GRAV  & 0.27       & 0.13 & 102.33            & 103\\
NUCL  & 27.20      & 9.69 & 18.85             & 56\\
Total   & 27.47      &  9.82 & 121.18           & 159\\
\hline
\end{tabular}
\caption{Energy extracted from the gravitational and nuclear reservoirs by different stellar populations. To convert the energies in Bethe (10$^{51}$ ergs) multiply the numbers by 1.6 10$^{-2}$.}
\label{tab:a}
\end{table}

In Table~\ref{tab:a}, the results for various populations are indicated. A first striking point is that
the low and intermediate mass stars differ from the massive stars by their ratio of $E_{\rm grav}$ to
$E_{\rm nucl}$. In the low and intermediate mass star regime, nuclear energy dominates being about 90 times
larger than the gravitational energy. For massive stars, the gravitational energy is about 
5.5 times larger than the nuclear energy. This comes of course from the fact that massive stars, at the end
of their lifetimes, create remnants that are much more compact. Globally, a stellar generation will
produce about twice as much energy by contraction as it produces through nuclear reactions.

Where does the bulk of gravitational energy extracted by the end of the
star's life goes? Is this just thermal photon emission, is it in neutrinos, or is it
in mechanical energy? We know the answer in the case of
massive stars which end in core collapse supernovae.
In that last case, most of the gravitational energy is released at the time of the supernova
explosion and is ejected as neutrinos (10$^{53}$ ergs). About 1\% is
released as kinetic energy (10$^{51}$ ergs) and about 1/10000 as radiation (10$^{49}$ ergs).
For small and intermediate mass stars, the energy from gravity is used to
compensate for the radiation losses at the surface during the lifetime of the
star. A part of it compensates for losses under the form of neutrino
emissions in the last evolutionary phases of these stars. 

For all these stars, low, intermediate and massive stars, most of the energy driving stellar winds comes from the nuclear
reservoir, since most of the time, stellar winds are driven by radiation during
the long nuclear burning phases.

%\section{The impact of nuclear power in stars}

\section{From nuclei to galaxies}

As written above, if gravity has the  leading role governing the evolution of stars, nuclear reactions have a deep impact on how this evolution occurs:
\begin{itemize}
\item They make stars live much longer than they would in case only the gravitational energy reservoir would be at their disposal.
\item They modify the chemical composition of stars, changing their thermodynamic properties, the equation of states and the opacity,
shaping the evolutionary tracks and the distribution of stars in the Hertzsprung-Russel diagram, the photometric and the chemical evolution of galaxies.
\item Thanks to nuclear reactions, stars build up the elements needed for the planet formation as well as for the apparition of their inhabitants (if any!)
triggering thus culture in the cosmos!
\end{itemize}

A universe without nuclear reactions would be completely different from the one we know and we would not be there to study it!
To conclude this brief paper, we would like to underline a few reactions whose impact is important in the frame of
topical astrophysical problems. For the sake of brevity, only a few comments will be made and references are given for more details. 

\subsection{$^{14}$N(p, $\gamma$)$^{15}$O and the age of globular clusters}

This reaction occurs in H-burning regions. It is responsible for the synthesis of nitrogen in the Universe.
The rate of $^{14}$N(p, $\gamma$)$^{15}$O from \citet{Muk} is about half the NACRE value (\citet{Angulo et al. 2009}) for temperatures below $10^{8}$ K, and compares well with other determinations like \citet{luna06}. In the low-mass domain, the effects of lowering this rate has been studied by \citet{imbr04} and \citet{weiss05}. They describe a slower H-burning process, and shallower temperature profiles leading to a more extended and slightly hotter core. The turn-off point is shifted towards higher luminosities. This leads to some revision of the ages of globular clusters (\citet{imbr04}) that are increased by 0.7-1 Gyr with respect to ages obtained from models using the reaction rate from NACRE.
In the intermediate-mass domain, the studies of \citet{herw06,weiss05} show that with slower rates, the MS evolution occurs at higher luminosities, and that later, the blue loops during core He burning get significantly shorter.

%Ages in the Universe are not easy to estimate, 
%and the number of methods not so many. There is the possibility to date meteorites from radioactive nuclei and sometimes a similar method can be used to stars (but here the process is much more uncertain since it is more difficult to have a precise idea of the number of radioactive elements that were present in the initial composition of stars), this can be used
%to date the solar system mainly. Another method
%is to use  the rate of expansion of Universe, this allows to link a given observed redshift with an age of the Universe. Finally observed properties of stars, position in the HR diagram,
%asteroseismic properties, activities, gyro chronology are also used. Nowadays, it becomes more and more important to be able to obtain reliable ages in order for instance
%to obtain a reliable chronology of the events marking the planet formation process or to be able to provide good estimates for the ages of young distant starbursts in galaxies and infer
%their recent star formation rate. This reaction....

\subsection{$^{12}$C($\alpha$, $\gamma$)$^{16}$O and the yields of carbon and oxygen}

This reaction occurs in He-burning regions of stars. It is responsible for the synthesis of carbon and oxygen in the Universe. It has also an impact on many outcome of stellar models,
the core He-burning lifetime, the formation of blue loops, the nature and properties of the stellar remnants. The rate from \citet{Kunz} is around 0.6-0.8 the NACRE value below $6\cdot10^{8}$ K, and around 1.1-1.4 the NACRE value above this temperature. Previous studies by \citet{WW} and \citet{imbr01} explored the effects of varying this rate. A stronger rate in the He-burning temperature range leads to larger cores, lower $^{12}$C and higher $^{16}$O yields.

\subsection{$^{17}$O($\alpha$, $\gamma$)$^{21}$Ne and the origin of s-process elements at low metallicity}

This reaction occurs in He-burning regions of stars. The ratio of this reaction with respect to the reaction $^{17}$O($\alpha$, n)$^{20}$Ne is
important to estimate the output of the $s$-process in massive stars. Let us explain why.
At the end of the core He-burning of massive stars, reactions like $^{22}$Ne($\alpha$,n)$^{25}$Mg
produce neutrons, which can then be captured by iron-peak nuclei to produce neutron-rich nuclei (the so-called weak s-process). 
It is well known that $^{16}$O is a neutron poison capturing them through the reaction $^{16}$O(n, $\gamma$)$^{17}$O, and thus
preventing them to form heavier nuclei. $^{16}$O is particularly relevant here since it is very abundant at the end of the core He-burning phase.
Now the neutron can be released again by $^{17}$O($\alpha$, n)$^{20}$Ne. Depending on this rate (and more precisely on how it compares with the competing channel $^{17}$O($\alpha$, $\gamma$)$^{21}$Ne), significative differences are obtained in the production of s-process elements
(see \citet{Frisch}). Before we discuss the present uncertainties of the reaction rate of $^{17}$O($\alpha$, $\gamma$)$^{21}$Ne, let us recall the broader context of s-process elements production in massive stars at very low metallicity.

The question of the s-process in massive stars has been recently rediscussed in the frame of rotating models. 
It was shown that in massive rotating stars at very low metallicity, the weak s-process could be much stronger that what was believed until now
(\citet{Pign}; \citet{Frisch}). In metal-poor rotating massive stars, much more
$^{22}$Ne is produced, boosting the s-process. 
This changes deeply the current view on how to interpret the presence of some of these s-process elements in very old stellar systems,
which acquired their present bulk chemical composition at such an early time in the history of the Universe that only massive stars
had time to contribute to their initial composition. However, the classical models of massive stars (with no rotation) produce a too weak s-process for accounting for the observed values.
With rotation, metal-poor  massive star models can produce s-process elements in much larger amounts than previously anticipated and they 
may explain the values of s-process elements observed in one of the oldest globular clusters (\citet{Chiappini et al. 2011}). But of course
to set on more quantitative grounds this conclusion, more accurate nuclear reaction rates are needed, in particular, the knowledge of the ratio of 
$^{17}$O($\alpha$, $\gamma$)-channel with respect to the $^{17}$O($\alpha$, n)-channel has to be improved.

% if the rate for the ratio of the rates for $^{17}$O($\alpha$, $\gamma$)$^{21}$Ne is much lower than the rate for $^{17}$O($\alpha$, n)$^{20}$Ne. When rates for $^{17}$O($\alpha$, $\gamma$)$^{21}$Ne and $^{17}$O($\alpha$, n)$^{20}$Ne from respectively Caughlan and Fowler (1988) and
%NACRE are used, at typical He-burning temperatures, the ($\alpha$, n)/($\alpha$, $\gamma$) ratio
% increases from 6.6 (at T8 = 2:5) to 16.5 (at T8 = 3:5).
%This shows that most of the neutrons captured by $^{16}$O are released
%again. But these neutrons can be recaptured by $^{16}$ which is a very abundant species at that stage, thus the reaction
%($\alpha$, $\gamma$) still play an important role.  

Typically, when rates for $^{17}$O($\alpha$, $\gamma$)$^{21}$Ne and $^{17}$O($\alpha$, n)$^{20}$Ne from respectively \citet{CF88} and
NACRE are used, at typical He-burning temperatures (between 250 and 350 $\times$ 10$^6$ K), the ($\alpha$, n)/($\alpha$, $\gamma$) ratios
are between 7 and 17. This may appear already as a serious advantage for the neutron releasing reaction but actually with such ratios, the
($\alpha$, $\gamma$) remain quite important because the neutrons can be recaptured by $^{16}$O! 
It happens that for the poisoning effect of $^{16}$O to disappear completely, a ratio about 1000-times lower should be obtained.
\citet{Descouvemont 1993} proposed a 1000-times lower ($\alpha$, $\gamma$)-channel with respect to the ($\alpha$, n)-channel. If this result holds true this
would have important consequences for the s-process element production. Some quantitative examples are given in \citet{Frisch}.

\subsection{$^{24}$Mg(p, $\gamma$)$^{25}$Al and the Mg-Al anticorrelation in globular clusters}

Globular clusters,  in contrast with previous common wisdom, were not formed from one
and unique gigantic starburst  more than 10 Gy ago. Instead, it appears that these systems hosted more than one star-formation episodes (see the review by \citet{Grat}). 
Let alone, this fact is not so much unexpected in view of the relatively large mass of these objects. What is however quite puzzling is the fact
that  the chemical composition
of stars of the generations that succeeded to the first one present very strange chemical patterns. Actually their chemical patterns reflect mainly, if not only, the expected pattern associated to hydrogen burning regions. How did the stars manage to expel only H-burning products? Different scenarios have been proposed (see the above review). One of the difficulty of many of these models however is to reproduce the correct extent of the Mg-Al anticorrelation. It appears that the use of the rate by
Powell et al (1999) for the reaction $^{24}$Mg(p, $\gamma$)$^{25}$Al produces a too small destruction of Mg and of course a too small production of Al to
reproduce the observed values. Thus, the question is whether this rate could be underestimated (see discussion in
\citet{Dec}).

Many more examples could have been chosen to illustrate the importance of having a better knowledge of nuclear physics. There is no doubt that
refining the estimates of some key nuclear reaction rates to within smaller error bars than presently available  will allow to make great progresses in many areas of astrophysics. 
The chemical abundances of the interstellar medium, reflected at the surface of non-evolved low mass stars, represent an archive of past star forming events and thus an access to the
histories of clusters and galaxies.

%%%%%%%%%%%%%%%%%%%%%%%%%%%%%%%%%%%%%%%%%%%%
%% SAMPLE TABLE
%%
%% Shows the use of \tablehead and \tablenote
%% macros
%%%%%%%%%%%%%%%%%%%%%%%%%%%%%%%%%%%%%%%%%%%%

%%%%%%%%%%%%%%%%%%%%%%%%%%%%%%%%%%%%%%%%%%%%%%%%
%% BACKMATTER
%%%%%%%%%%%%%%%%%%%%%%%%%%%%%%%%%%%%%%%%%%%%%%%%





\end{document}